\documentclass{ws-procs9x6}

\def\squig{\sim\!\!}
\def\lesssim{\mathrel{\hbox{\rlap{\hbox{\lower4pt\hbox{$\sim$}}}\hbox{$<$}}}}
\def\gtrsim{\mathrel{\hbox{\rlap{\hbox{\lower4pt\hbox{$\sim$}}}\hbox{$>$}}}}

\begin{document}

\title{CENSORS: The VLT/VLA mJy Source Survey for High Redshift AGN Evolution}

\author{M.~H. Brookes and P.~N. Best}

\address{Institute for Astronomy,\\
Blackford Hill, \\
Edinburgh,\\ 
EH9 3HJ, UK\\ 
E-mail: mhb@roe.ac.uk, pnb@roe.ac.uk}


\maketitle

\abstracts{The Combined EIS-NVSS Survey Of Radio Sources (CENSORS) has been produced with the primary goal of investigating the cosmological evolution of the radio luminosity function. This 1.4GHz sample, complete to the $7.2$mJy level, contains 150 radio sources. Host galaxies are almost entirely identified in optical and near-IR bands and the sample is now approaching 70\% spectroscopic completeness. We show preliminary results demonstrating how CENSORS will improve upon previous work and how it is applicable to other projects.}

\section{Introduction}

Radio galaxies and radio loud quasars are among the most powerful objects
in the Universe. Observable at high redshifts, they trace large scale
structure and are associated with the most massive black holes[\refcite{rad_bhmass}], they therefore have great potential for revealing the evolution of massive
galaxies and investigating the relation between active galactic nuclei
(AGN) and galaxy formation. 
However there are still many questions facing radio-loud AGN astrophysicists.
For example, the high redshift cosmological evolution of strongly radio active sources is a property that has not been resolved and so provides a route by which we may gain a better understanding of these objects. This is of far-reaching importance: given their association with the most massive black holes, we would also learn about the evolution of the upper part of the black hole mass distribution.

In 1990 Dunlop and Peacock[\refcite{dp90}] undertook a  study of
cosmic evolution of the radio luminosity function (RLF) presenting the first evidence of a decline
in the comoving number density 
of powerful radio sources beyond z $\squig$ ~2.5 (the redshift cut-off). 
Since then numerous advances have provided a good consensus in
the determination of the low redshift RLF (eg.[\refcite{mob1999}]), but the high redshift evolution and the reality of the redshift cut-off in the radio source population remain areas of controversy. 
For example, the deep sample of Waddington et al.[\refcite{wad2001}] shows evidence of a cut-off in number density of low luminosity sources beyond z $>$ 2, but has insufficient sky coverage to investigate the most luminous sources. Conversely, the sample of Jarvis et al.[\refcite{jarvis2001}] proved too shallow to resolve the issue.
 Shaver et al.[\refcite{shaver}] claimed evidence for a sharp decline in number density of flat spectrum sources between $z \sim 2.5$ and $z \sim 5$. However Jarvis and Rawlings[\refcite{jar_raw}] showed that this result was erroneous due to a failure to properly account for the  spectral index of the sources.
So high redshift RLF evolution of radio sources remains ill-understood.

\section{CENSORS}
\label{define}
\label{results}
\begin{figure}[h]
\centerline{\epsfxsize=6.7cm\epsfbox{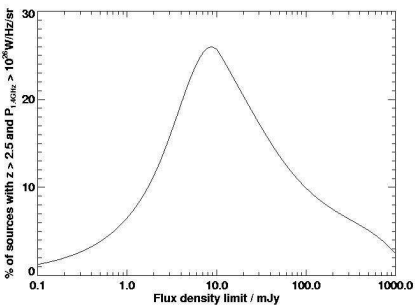}}   
\caption{This plot, based upon the pure luminosity evolution models of Dunlop and Peacock (1990), demonstrates that a 1.4GHz radio survey with flux limit just below 10mJy will maximise the information on high redshift powerful radio sources.\label{why_best}}
\end{figure}

Controversy in earlier work begins at z $\squig$ 2.5, so CENSORS was designed to maximise information at these redshifts (see Figure \ref{why_best}). 
CENSORS contains all the NVSS sources above $7.2$mJy that are within the ESO Imaging Survey (EIS) Patch D ($2\times3$ degrees; $09\ 51\ 36\ -21\ 00\ 00$). High resolution VLA imaging of CENSORS allowed
EIS I band imaging (I $\lesssim$ 23) to be used to
identify host galaxies using a maximum likelihood method
analysis[\refcite{censors1}]. Remaining sources
have been identified using K band imaging using UKIRT, the AAT and the VLT[\refcite{brookes1}].
Following spectroscopy, at the VLT, the ESO 3.6m and the AAT[\refcite{brookes2}], the sample is also now
approaching 70\% spectroscopic completeness with further follow-up on-going.

\section{Results}
\label{results}

\begin{figure}[h]
\centerline{\epsfxsize=6.7cm\epsfbox{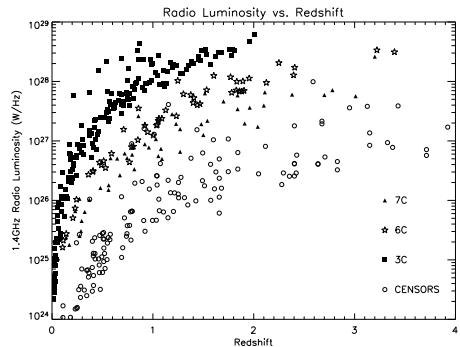}}   
\caption{This P-z plane shows how adding CENSORS to samples like 3C, 6C and 7C will help to break the radio luminosity-redshift degeneracy.\label{pz}}
\end{figure}

\begin{figure}[h]
\centerline{\epsfxsize=12cm\epsfbox{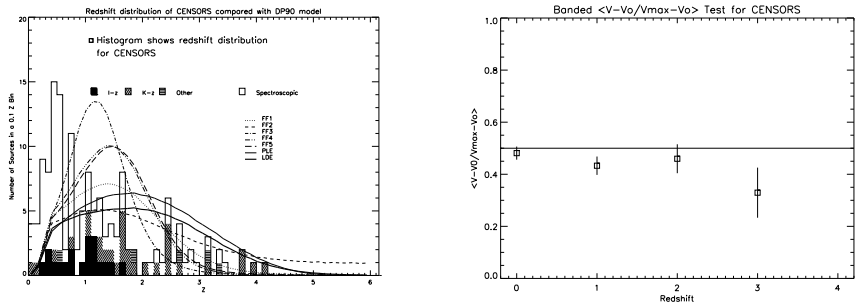}}   
\caption{{\bf {\it Left}}:The redshift distribution of the CENSORS sample compared with the model predictions of Dunlop and Peacock 1990. FF indicates the free form models.{\bf {\it Right}}: A banded ${\frac{V}{V_{max}}}$ test for CENSORS, shows negative evolution at the highest redshifts. \label{zdist}}
\end{figure}

Given the current status of CENSORS we may establish its potential to build upon previous work using the spectroscopic redshifts where we have them, and estimated reshifts using the K-z and I-z relations where we don't.
For example, the additional coverage it offers in the radio luminosity vs.~redshift plane is illustrated by Figure \ref{pz}.
By plotting the estimated redshift distribution (Figure \ref{zdist}) we demonstrate that CENSORS improves upon the predictions of Dunlop and Peacock, as some of their allowed fits clearly depart from the CENSORS distribution. Hence, we are probing information which was not included in the former investigation. We also show a banded ${\frac{V}{V_{max}}}$ test for CENSORS. In the highest redshift bin ${\frac{V}{V_{max}}}$ falls well below $0.5$. Although the error bars here do not account for the estimation in the redshift distribution, this gives tentative indications of negative evolution at high redshift.

To clarify the high redshift evolution we intend to use this sample, in conjunction with others, such as 3CRR[\refcite{lrl}], BLR[\refcite{blr}] and the Parkes selected regions[\refcite{psr}], to model the evolution of the RLF. Not only will we improve on previous work by our choice of sample but, in cases where a spectroscopic redshift has not yet been obtained, we will model target redshifts with a photometric redshift probablitiy distribution. We will, therefore, be able to account for this uncertainty within the model and in the reliability of the results.

\section{Further Investigations}
CENSORS will also be useful in other important investigations. For example the nature of the K-z relation for radio galaxies and its dependence on radio luminosity, the comparison of evolution of radio loud and quiet sources and the dependence of sources properties on environment.

\end{document}